# Non-equilibrium dynamics in geometrically frustrated spin glass $Bi_2Fe_3GaO_9$ with a Cairo lattice

Desislava Mihaylova [1,2], Xinglong Chen [1], Daniel Phelan [1], Stephan Rosenkranz [1], Yu Li [1]

[1] Materials Science Division, Argonne National Laboratory, Lemont, Illinois 60439, USA

[2] Department of Physics, University of Washington, Seattle, Washington 98105, USA

We have explored the relaxation process of the spin glass phase in $Bi_2Fe_3GaO_9$, a geometrically frustrated magnet with a unique structure consisting of pentagonal building blocks known as the Cairo lattice. Using dc magnetization measurements on single crystals, we estimate the relaxation time across various temperatures and fields. Our results indicate that the relaxation time of $Bi_2Fe_3GaO_9$ follows the Arrhenius law as a function of temperature but remains relatively constant as the applied magnetic field varies. Further, through a carefully designed protocol, we observe significant rejuvenation and memory effects. Remarkably, the memory effects in $Bi_2Fe_3GaO_9$ are found to be more similar to those observed in classical spin glasses rather than in spin jamming systems, both in terms of the magnitude of the memory effect and the value of the exponent that describes the relaxation behavior. Our findings suggest that $Bi_2Fe_3GaO_9$ provides an excellent platform for investigating the time-dependent evolution of underlying magnetism using neutron scattering, paving the way for future research.

## 1. Introduction

The fundamental ideas behind spin glasses have captured the interests of scientists across diverse fields for decades, including physics [1-4], mathematics [5], biology [6], computer science [7], and psychology. As a prototype of complex systems in condensed matter physics, spin glasses exhibit fascinating phase transitions and exotic dynamics. These materials are characterized by randomly oriented magnetic moments that are frozen below a specific transition temperature, $T_g$. This behavior is a result of both quenched disorder and frustration, generating a rugged free energy landscape [8, 9] where systems can be trapped in metastable states separated by significant energy barriers. This ultimately leads to ergodicity breaking.

Positioned between complete randomness and perfect order, spin glasses pose great challenges for comprehensive description and theoretical understanding due to their highly irregular magnetic structure and substantial entropy of information at low temperatures. A mean-field solution for the infinite-ranged spin glass model introduced the concept of replica-symmetry breaking (RSB) [10, 11], successfully explaining the hierarchical energy scales and the ultrametric structure of phase space. Meanwhile, alternative theories such as the droplet model [12-14] offer different perspectives. Despite these theoretical advances, a significant gap still exists between theoretical predictions, computational simulations, and experimental observations. In order to bridge this gap, experimental investigations on various materials are required to probe the correlations over different length and time scales.

In the classical dilute metallic spin glasses like CuMn [15, 16], disorder is the leading factor for generating magnetic frustration via distance-dependent Ruderman-Kittel-Kasuya-Yosida (RKKY) interaction. In contrast, geometric frustration can emerge in perfect lattices without disorder, due to mismatches between neighboring antiferromagnetic interactions and lattice geometry [17, 18]. This has stimulated extensive research [19, 20] into geometrically frustrated materials, which can exhibit behaviors dramatically different from classical spin glasses, necessitating new theories and concepts to explain these exotic phases and dynamics [21-28].

Among the extensively studied geometrically frustrated materials, $Bi_2Fe_4O_9$ is unique as its $Fe^{3+}$ spins crystallize in a close representation of the Cairo lattice composed of irregular pentagons. Geometric frustration plays a pivotal role in determining its magnetic properties [29, 30]. While this compound exhibits long-range antiferromagnetic order below 240 K, partial substitution of Fe with non-magnetic Ga suppresses this order and induces a spin glass transition [31, 32], as depicted in Fig.1(b). Our recent neutron scattering experiments [33] reveal short-range magnetic correlations in Ga-doped $Bi_2Fe_4O_9$, resembling a theoretically predicted configuration on an ideal Cairo lattice [30]. This highlights the intricate relationship between magnetic behaviors and lattice geometry. However, the non-equilibrium dynamics of the spin glass state remain underexplored. Investigating these dynamics and their relation to geometric frustration is both intriguing and crucial.

In this manuscript, we present the first experimental investigation of the time-dependent glassy dynamics of single crystals of $Bi_2Fe_3GaO_9$ using dc magnetization measurements. This material undergoes a spin glass transition at approximately 30 K. In contrast to the parent $Bi_2Fe_4O_9$, where magnetic moments lie in the plane of the Cairo lattice [34], $Bi_2Fe_3GaO_9$ displays nearly identical magnetic susceptibilities for both out-of-plane and in-plane directions, suggesting isotropic magnetic behavior. Our measurements show a linear magnetic response to the applied field, implying the system remains far from saturation or magnetic instability within our measurement range up to H = 7 T. We explore the glassy dynamics via time-dependent dc magnetization measurements after thermal quenching in zero field and estimate the relaxation times using the Kohlrausch-Williams-Watts (stretched exponential) function [35-38]. Our findings reveal that the relaxation time remains independent of the applied magnetic field but varies monotonously with temperature, following an Arrhenius law. Furthermore, the relaxation time is also affected by the wait time of the sample prior to each measurement. Additionally, we examine rejuvenation and memory effects by introducing small temperature changes. Intriguingly, our results suggest that the memory effect, characterized by the ratio, $\frac{\chi_{dc} - \chi_{ZFC}}{\chi_{ZFC}}$, is significant compared to that in spin jamming systems and more reminiscent of classical spin glasses. Fitting the relaxation process with a stretched exponential function yields an exponent n ~ 0.39, close to that of dilute magnetic alloys with hierarchical and fractal energy landscapes. Given its dense population of magnetic moments and suitable relaxation timescales, we propose that $Bi_2Fe_3GaO_9$ offers a promising opportunity for investigating non-equilibrium dynamics in spin glasses using neutron scattering.

## 2. Methods

We have successfully grown high-quality single crystals of $Bi_2Fe_{4-x}Ga_xO_9$ with x = 1 using a combination of the flux method and top-seeded solution growth technique. In this process, a mixture of $Bi_2O_3$, $Fe_2O_3$, and $Ga_2O_3$ with a mole ratio of 73%, 20.25%, and 6.75%, respectively, was thoroughly mixed, placed in a platinum crucible, and transferred to a programmable, vertical-tube furnace. The mixture was heated at a rate of 50 K/hour until melted, and then held at that temperature for 20-40 hours to obtain a homogeneous solution. To facilitate the formation and separation of as-grown crystals, a platinum wire attached to an alumina rod tip was dipped into the center of the solution surface. Afterwards, the solution was cooled at a rate of 5-10 K/day until spontaneously nucleated crystals around the platinum wire were observed. It was then further cooled at a rate of 3-5 K/day for 10-20 days. The as-grown crystals were slowly pulled out of the solution which was finally cooled to room temperature at a rate of 50 K/hour. The resulting crystals exhibit well-developed facets, are black in color, and are several centimeters in size.

The orientation of single crystals was determined using X-ray diffraction on a Bruker D8 diffractometer equipped with APEX2 area detector and Mo K$\alpha$ radiation ($\lambda = 0.71073$ Å). Temperature- and field-dependent dc magnetization measurements were performed using a Quantum Design SQUID Magnetometer (MPMS 3). To conduct our experiments, we attached the samples on a quartz sample holder with a desired orientation using Loon Outdoors UV Knot sense. For time-dependent measurements, the sample was quenched to the target temperature at a rate of 35 K/min at zero field, followed by the application of a small magnetic field for the dc magnetization measurement. The wait time, $t_w$, refers to the period between temperature stabilization and the application of the small magnetic field. For temperature-dependent measurements of the memory effect, we used a step-cooling protocol in which the sample was rapidly cooled to a set temperature, $T_{aging} < T_g$, at a rate of 35 K/min at zero field. After being held at $T_{aging}$ for a period of time, the sample was further cooled to the base temperature. Then, the dc magnetic susceptibility measurement was performed as a function of temperature upon warming. The memory effect is manifested as a shoulder at $T_{aging}$ in the measured data.

### 3. Experimental Results

We present in Figure 1(c) and 1(d) the temperature-dependent behavior of the dc magnetic susceptibility measured following both zero-field-cooling (ZFC) and field-cooling (FC) procedures, with the magnetic field oriented either parallel or perpendicular to the c-axis. A noticeable bifurcation between the two curves can be seen below the spin glass transition temperature. A Curie-Weiss fit to this data from 100 K to 300 K yields consistent parameters: $T_{weiss} = -53\ K$ and $\mu_{eff} = 2.6\ \mu_B$. These values are significantly lower than previously reported for polycrystalline samples [31]. Interestingly, the observed spin glass transition temperature, $T_g \sim 30$ K, is higher than that of polycrystalline $Bi_2Fe_{4-x}Ga_xO_9$ with x = 1 [31, 32]. This discrepancy may arise from the actual Ga concentration in our single crystals being lower than the nominal mole ratio of the initial solution. Our magnetization measurements also reveal no visible magnetic anisotropy, in sharp contrast with the in-plane magnetic moments of the parent $Bi_2Fe_4O_9$ [34, 39, 40]. However, the absence of long-range magnetic order suggests that our sample's Ga concentration might be near the top or slightly right side of the spin glass dome, as shown in Figure 1(b).

Despite these findings, the estimated Weiss temperature is significantly higher than the spin glass transition temperature, implying the presence of magnetic frustration [17, 29]. Figures 1(e) and 1(f) show that the field-dependent magnetization is nearly linear with minimal hysteresis within the field range of our measurements. The magnitude of magnetization in the whole range of measurements is significantly lower than saturation, suggesting strong antiferromagnetic interactions.

To understand the intrinsic glassy dynamics, we explored the time-dependent behavior of magnetization in this crystal. Figure 2(a) outlines the protocol for dc magnetic susceptibility measurements [41, 42, 43]. We define t = 0 when the sample reaches the desired temperature. The uncertainty for this is estimated to be less than one minute, due to varying times for the application of magnetic field and the measurement for the first data point. Figures 2(b-e) present time-dependent data measured at 12 K with a 100 Oe applied field, fitted with various exponential functions to estimate relaxation time. These include a stretched exponential function, $M(t) = M_0 - A \times e^{-(t/\tau)^n}$ [35-38, 41, 44]; a regular exponential function, $M(t) = M_0 - A \times e^{-t/\tau}$ ; the sum of two exponential functions, $M(t) = M_0 - A_1 e^{-t/\tau_1} - A_2 e^{-t/\tau_2}$; and an alternate exponential function [45, 46], $M(t) = M_0 - At^{-b} e^{-(t/\tau)^n}$. As is clearly seen from Figure 2, the stretched exponential function provides the best fit and is selected for further analysis.

In Figure 3, we summarize the relaxation time, $\tau$, estimated at various temperatures and fields. Figure 3 (a-d) demonstrates a monotonous increase in relaxation time as the temperature decreases, suggesting potential divergence at lower temperatures, consistent with the Arrhenius law, $\tau = \tau_0 e^{(E_b/(k_B T))}$. The solid curves are guides for the eye rather than precise fits. In some glassy systems, the relaxation time diverges at finite temperatures, depicted by the Vogel-Fulcher (VF) law [1], $\tau = \tau_0 e^{(B/(T-T_0))}$. However, our data are insufficient to clearly distinguish between these scenarios, and our attempts to fit with the VF law resulted in a negative $T_0$.

We also investigate the effect of magnetic field on relaxation time, with results for T = 15 K and 18 K shown in Figure 3(e) and 3(f). These results suggest that the magnetic field applied for dc magnetization measurements does not significantly influence the relaxation time, though the energy landscape of the system could be potentially modified.

Generally, the observed relaxation time reflects the evolution of the magnetic structures over time. One intuitive explanation involves the growth of hypothetical clusters. Following this hypothesis, we let the sample age for a period of time, defined as wait time, after temperature stabilization and before applying magnetic fields, as illustrated in Figure 4(a). Figure 4(b) and 4(c) present the measured relaxation time of $Bi_2Fe_3GaO_9$ as a function of wait time at T = 15 K and 18 K. The relaxation time increases with longer wait times, in agreement with the cluster growth model, where larger clusters take longer to flip between different states.

In figure 5, we present the fitted exponent values from the stretched exponential function obtained by fitting various data. Despite slight variations, the fitted exponent values remain consistently around 0.39 across different fields and temperatures. While this exponent value is close to that obtained in dilute magnetic alloys, it is far from the value of 0.6 in most densely populated magnets [48]. It is worth noting that fixing the exponent for fitting all the data leads to consistent results for temperature, field, and wait time dependencies. The estimated timescales vary from minutes to tens of minutes across various temperatures, falling within an appropriate time window for many experimental techniques. These results suggest that $Bi_2Fe_3GaO_9$ is an ideal material for studying time-dependent glassy behavior.

Besides slow relaxation of magnetization, spin glasses are known for intriguing rejuvenation and memory effects, studied extensively via ac magnetic susceptibility measurements in previous literature. A typical measurement comprises of three stages: aging, rejuvenation and memory effects, each separated by small temperature changes [43]. In the first stage, as a spin glass is quenched from above $T_g$, it becomes trapped in a metastable state and slowly relaxes toward a hypothetical equilibrium. In the second stage, a small change in temperature induces completely refreshed relaxation dynamics, resembling those in an as-quenched sample – a rejuvenation effect. Finally, when the sample temperature returns to its original value, the resultant relaxation behavior connects continuously to the first stage's curve, demonstrating a memory of the aged state [43].

In Figure 6 (b-e), we present dc magnetization measurements following a protocol shown in Figure 6(a). We employed both positive and negative thermal cycles with $\Delta T = \pm 2$ K, as shown in Figure 6(b) and 6(d). Unlike ac susceptibility measurements, the presence of a static magnetic field could affect our interpretation of the measured data in two ways. First, it introduces an average magnetization independent of time within each stage. To emphasize time-dependent components, we calculate the derivative, $d\chi_{dc}/dt$, and present the results in Figure 6(c) and 6(e). We argue that $d\chi_{dc}/dt$ serves the same role as $\chi_{ac}$ in previous ac susceptibility measurements. Second, a finite magnetic field during cooling could create distinct magnetic states due to ZFC and FC susceptibility bifurcation, as illustrated in Figure 6(f).

In the positive thermal cycle in Figure 6 (b), raising the sample temperature from $T_M = 20$ K to $T_M + \Delta T = 22$ K reveals a rejuvenation effect possibly related to a redistribution of magnetic configurations. However, when the temperature returns to $T_M$, the sample undergoes a field-cooling procedure, entering a state different from its original one in stage 1. Therefore, a memory effect is absent. Conversely, during the negative thermal cycle in Figure 6(d), the sample follows a field-cooling procedure first in stage 2 and then recovers the initial state in stage 3, exhibiting a memory effect, as shown in Figure 6(e).

To further confirm the memory effect in $Bi_2Fe_3GaO_9$, we employed a step-cooling procedure and measured *dc* magnetic susceptibility as temperature increases, as illustrated in Figure 7 (a). The sample was first cooled in zero field to $T_{aging} = 20$ K and held there for a period ranging from 5 minutes to 10 hours before being further cooled to base temperature. A magnetic field of 100 Oe was then applied, and measurements were conducted upon warming. As seen in Figure 7(b), a significant memory effect is evident as a flattening of the temperature-dependent susceptibility when approaching $T_{aging}$. Using the ZFC curve as a reference, we calculated the normalized changes of magnetic susceptibility according to $(\chi_{dc} - \chi_{ZFC})/\chi_{ZFC}$, as shown in Figure 7(c). Despite the absence of a dip in the temperature-dependent susceptibility, the normalized magnitude of the memory effect in Figure 7(c) is the same as that observed in classical spin glasses CuMn2% and CuMn45%, and significantly larger than in spin jamming systems [47, 48]. These results, combined with the fitted exponent in Figure 5, strongly suggest that the energy landscape of $Bi_2Fe_3GaO_9$ is characterized by a hierarchical rugged structure instead of a nearly flat bottom [49], a prominent feature of spin glasses. Furthermore, in Figure 7(c), the depth of the valley changes rapidly within the first few minutes, and then gradually approaches a potential saturation point, consistent with the stretched exponential decay of magnetization over time.

## 4. Discussion and Conclusion

Although theories about spin glasses are often complex and still in discussion, an intuitive picture for understanding the observed experimental phenomena can still be helpful. Our recent diffuse neutron scattering investigations have revealed that the magnetic diffuse scattering in $Bi_2Fe_3GaO_9$ may be considered as a kinetic process of cluster growth, inherently associated with the aging behavior. However, this straightforward explanation is insufficient to account for the rejuvenation effect. Advanced concepts, such as temperature chaos, could provide insights, implying the existence of numerous, uncorrelated equilibrium magnetic states at different temperatures. The coexistence of rejuvenation and memory in spin glasses further complicates the scenario, since any small temperature perturbation could induce temperature chaos at all length scales, ~~which~~ unavoidably destroying the memory of the past state in spin glasses. Some theories suggest that there could exist low-lying sponge-like excitations [43, 50] across several length scales and might provide a promising explanation. While it remains uncertain whether such structures exist in spin glasses, it is crucial to develop new experimental techniques to detect these underlying magnetic configurations.

Even if temperature chaos is present in spin glasses, its effect can be constrained by an overlap length, which reflects the degree of correlation between two magnetic states before and after a temperature change and depends on both the amplitude of temperature change and aging time. Since the valley in Figure 7(c) represents the residual memory, it might serve as an indicator of the overlapping between the two magnetic states before and after a temperature change. Therefore, to ensure continuous measurements of aging dynamics without interruption, the temperature variation has to be controlled within the temperature range at the bottom of the valley. This sets an upper limit of the tolerance for temperature fluctuations during an experiment. Specifically, to monitor the time-dependent evolution of magnetic structures in $Bi_2Fe_3GaO_9$, temperature fluctuations must be held below 1 K, which is easily achievable with the existing sample environment.

In conclusion, we have investigated the non-equilibrium dynamics of the geometrically frustrated spin glass $Bi_2Fe_3GaO_9$, hosting a Cairo pentagonal lattice. We explored the slow relaxation process over time below the glass transition temperature and measured its response to external factors such as temperature, magnetic field, and sample history. Notably, the relaxation time in $Bi_2Fe_3GaO_9$ is not dependent on applied magnetic fields, but increases as temperature drops, following the Arrhenius law. We also explored the effects of wait time, as well as the rejuvenation and memory effects. While geometric frustration appears to contribute to the formation of the spin glass phase in $Bi_2Fe_{4-x}Ga_xO_9$, the magnitude of memory effect does not significantly differ from that of traditional spin glasses. We speculate that the observed spin glass dynamics may be related to the growth of speculated clusters or domains with unconventional surfaces. With its dense population of magnetic moments and a suitable time window, $Bi_2Fe_3GaO_9$ serves as an ideal platform for investigating the time-dependent dynamics in spin glasses through neutron scattering. Our characterization of the relaxation time of $Bi_2Fe_3GaO_9$ at various temperatures and fields provides valuable guidance for further research.

## Acknowledgements

This work was supported by the U.S. Department of Energy, Office of Science, Basic Energy Sciences, Materials Sciences and Engineering Division.

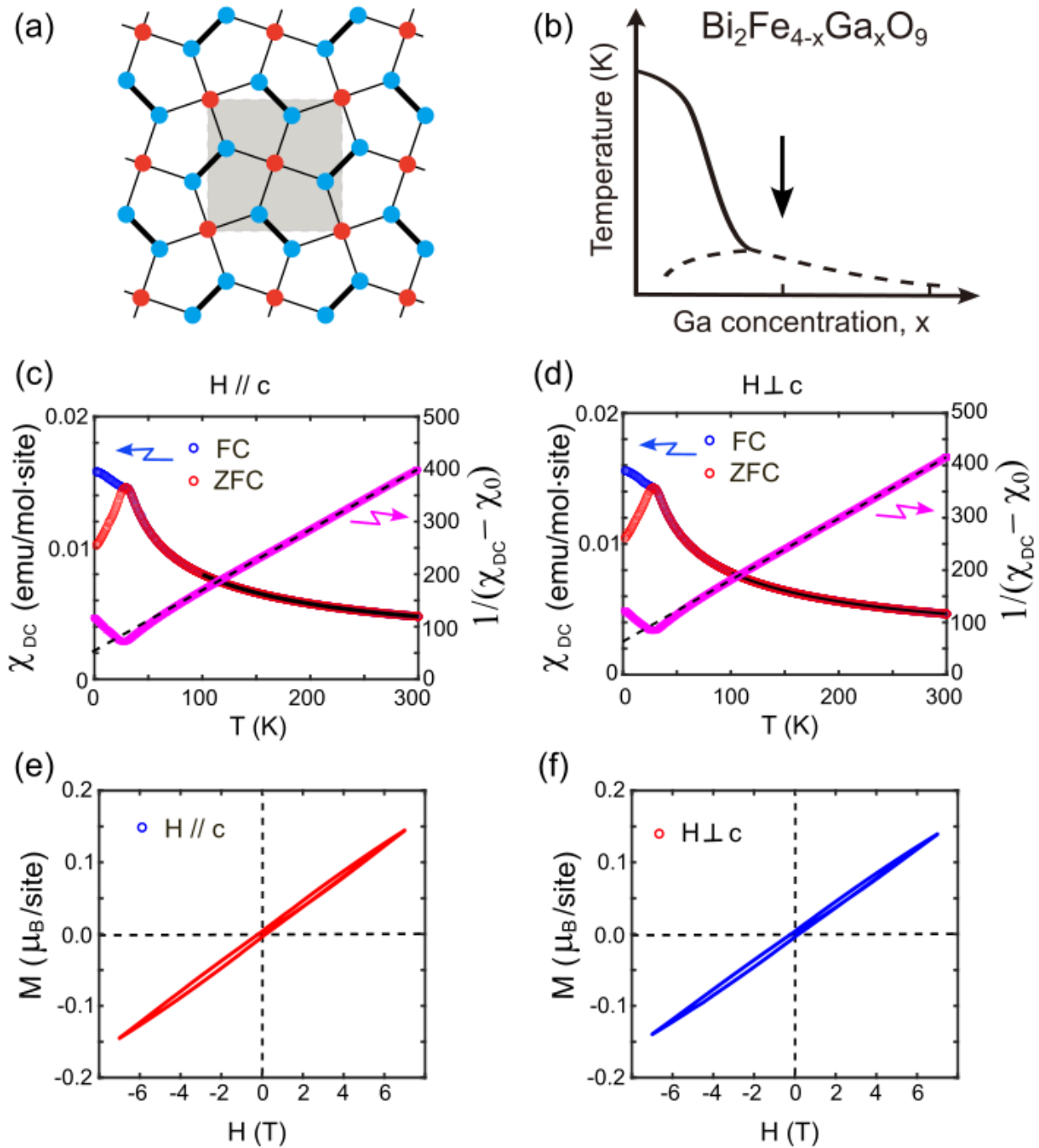

**Figure.1** (a) Illustration of a 2-dimensional Cairo lattice with two different sites. (b) Schematics of the magnetic phase diagram [31,32] of $Bi_2Fe_{4-x}Ga_xO_9$. (c) and (d) The dc magnetic susceptibility, M/H, as a function of temperature after zero-field-cooling (ZFC) and field-cooling (FC) procedures, with the magnetic field, H = 1000 Oe, oriented along out-of-plane and in-plane directions. The corresponding inverse susceptibilities is also presented with their values on the right-side axis. (e) and (f) The magnetic hysteresis measured at 1.8 K with the applied field either parallel or perpendicular to the c-axis.

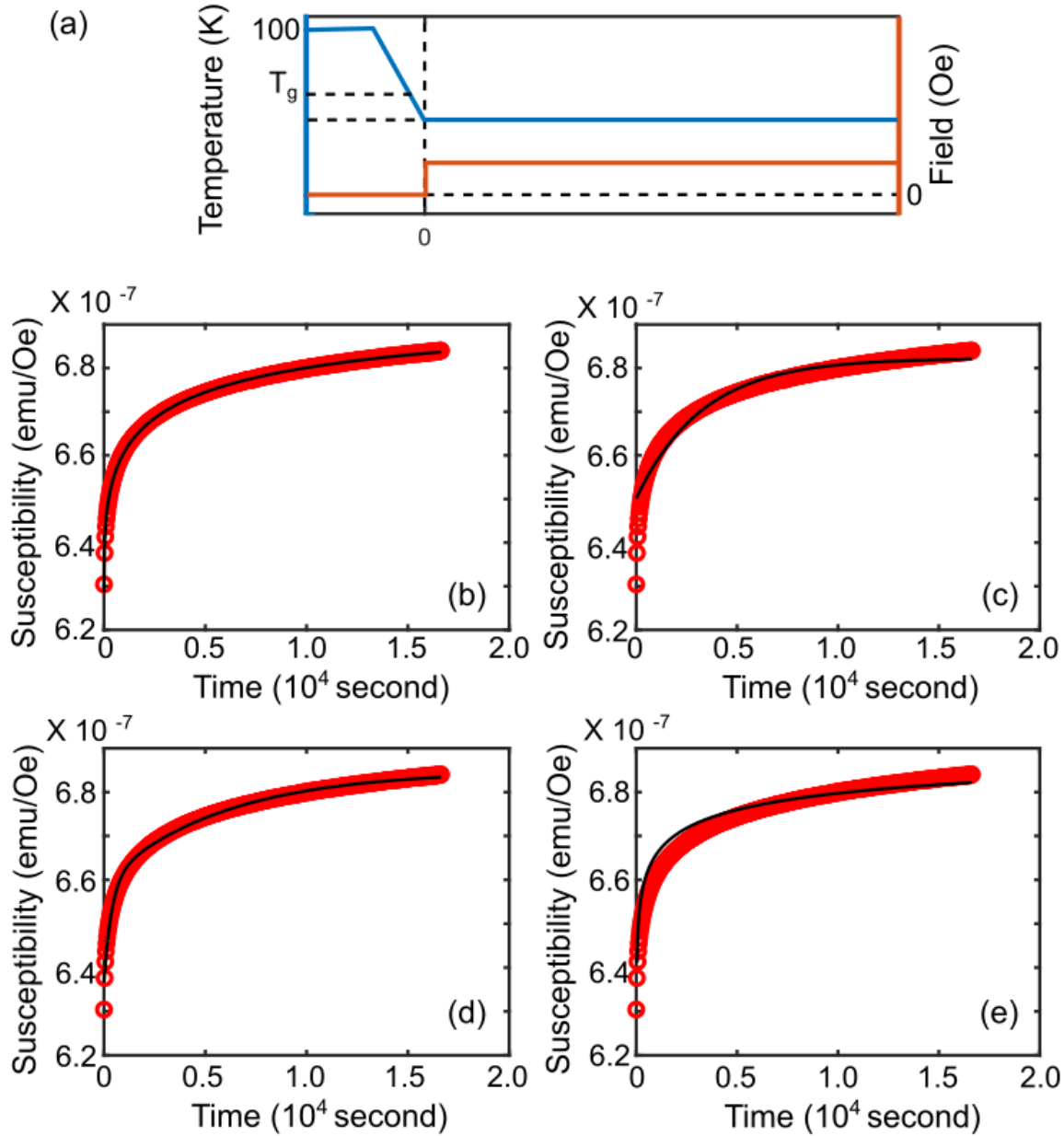

**Figure 2**. (a) Illustration of the field and temperature procedure for the measurement of the relaxation behavior. (b-e) the dc magnetic susceptibility, M/H, as a function of time, measured at T = 12 K with a dc field of 100 Oe. The wait time, as defined ~~later~~ in the main text, is zero for this measurement. The black curves are the fit result of the same experimental data with different fitting functions, namely, a free stretched exponential function [35, 41, 36, 44, 37, 38] in (b), regular exponential function in (c), sum of two exponential functions in (d) and an alternate stretched exponential function [45, 46] in (e). It is apparent that the free stretched exponential function gives the best fitting result.

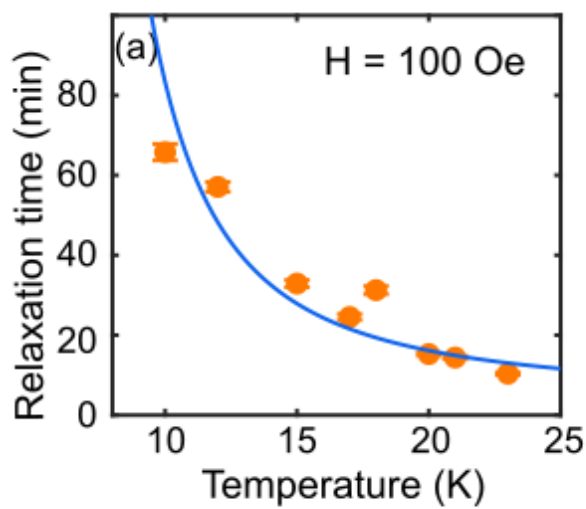
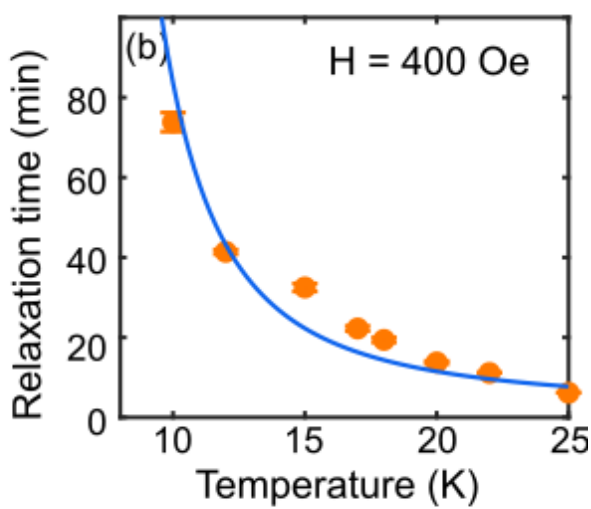
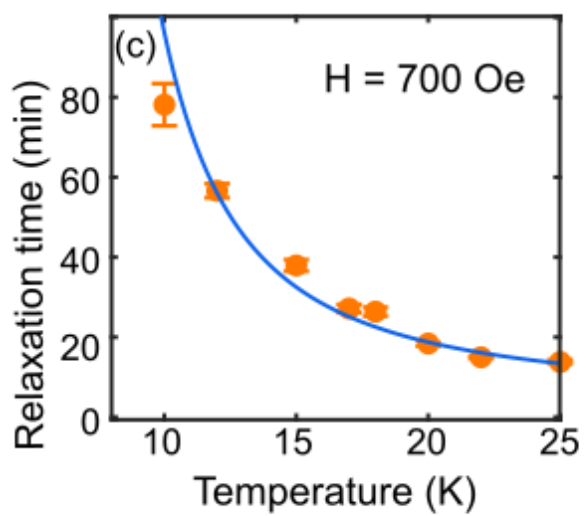
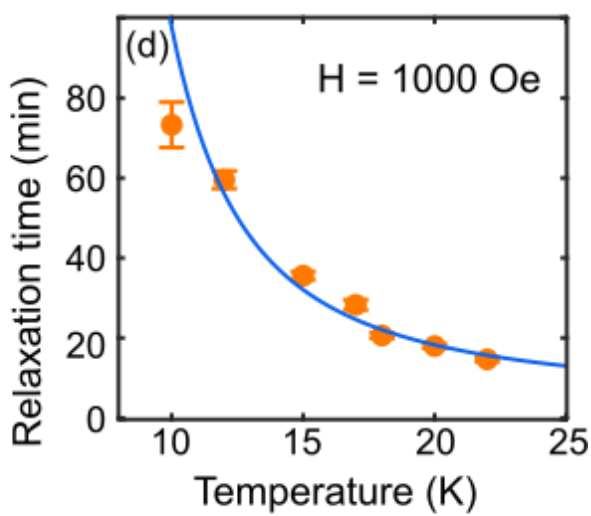
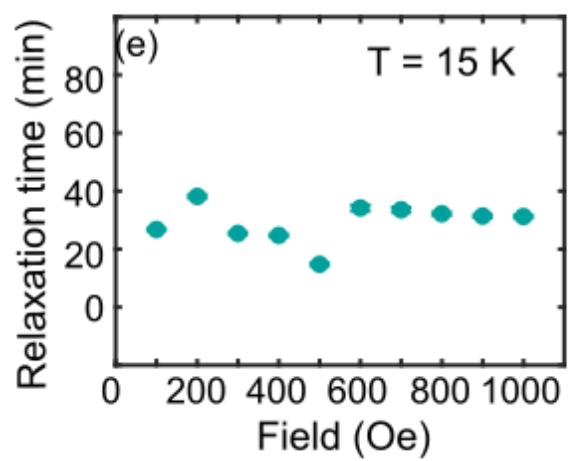
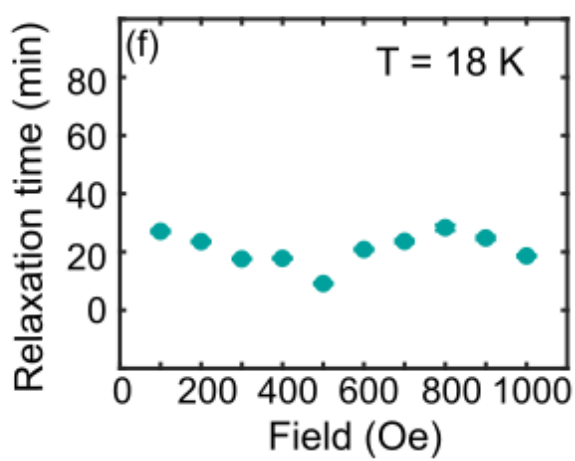

**Figure 3**. (a-d) The relaxation time estimated by fitting with a free stretched exponential function, as a function of temperature. The applied magnetic fields are H = 100 Oe, 400 Oe, 700 Oe, and 1000 Oe. The solid blue curve is a guide for the eyes following an Arrhenius equation. (e) and (f) The estimated relaxation time as a function of applied field at T = 15 K and 18 K, respectively.

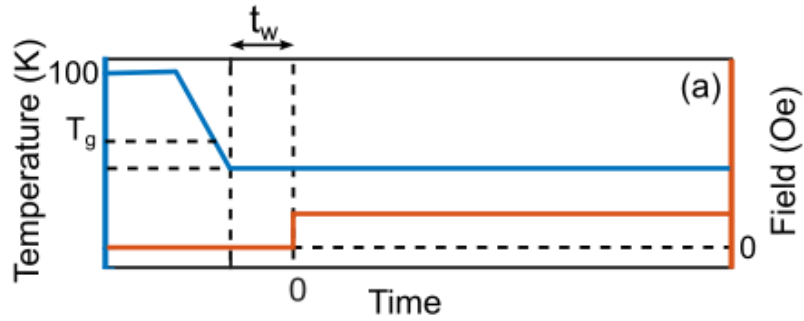

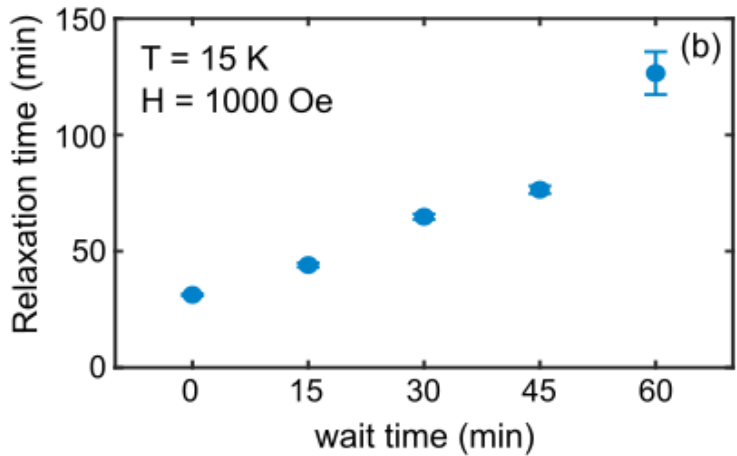

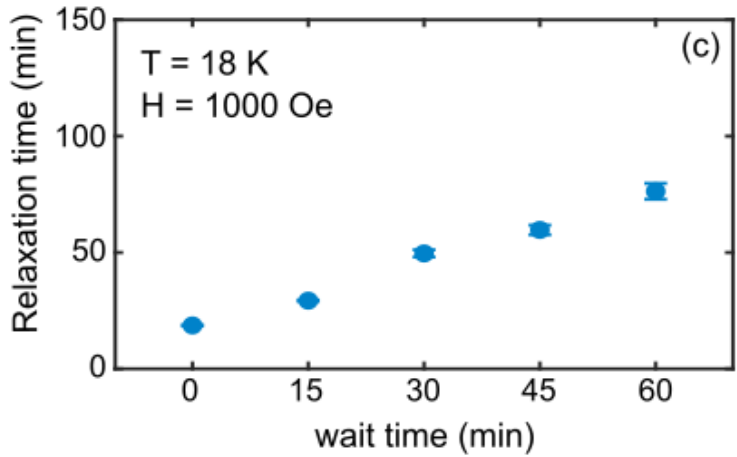

**Figure 4.** (a) Illustration of the field and temperature procedure for the following measurements. A wait time was set after the stabilization of the sample temperature and before the starting of measurements. (b) and (c) The estimated relaxation time varies with wait time at T = 15 K and 18 K respectively.

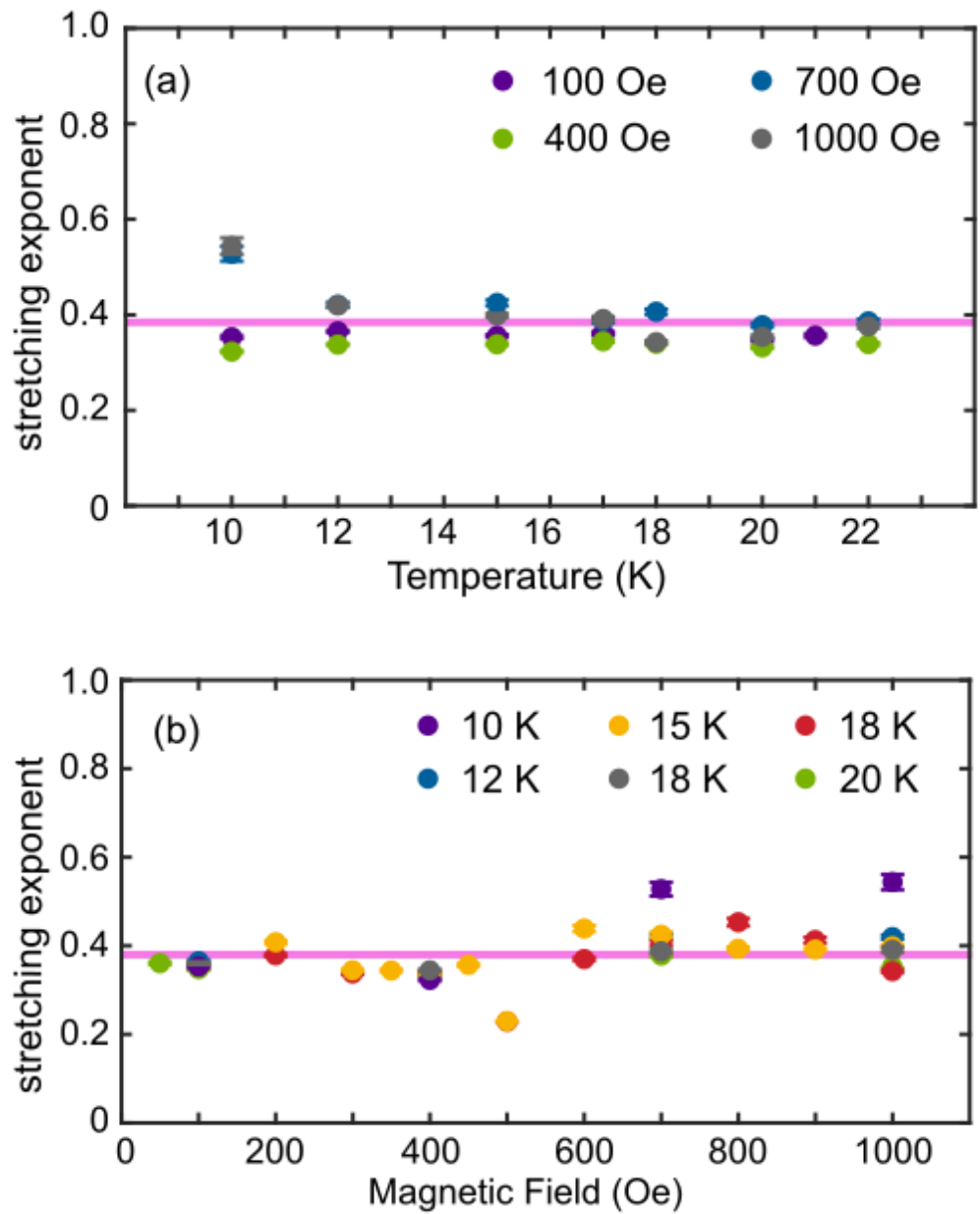

**Figure 5.** The fitted values of the stretching exponent, $n$, in the free stretched exponential function, $\chi_{dc} = \chi_0 - A e^{-(t/\tau)^n}$ for measurements at a series of temperatures and applied fields. These data are plotted as a function of either temperature in (a) or field in (b). The pink line is a guide to the eyes at n = 0.39.

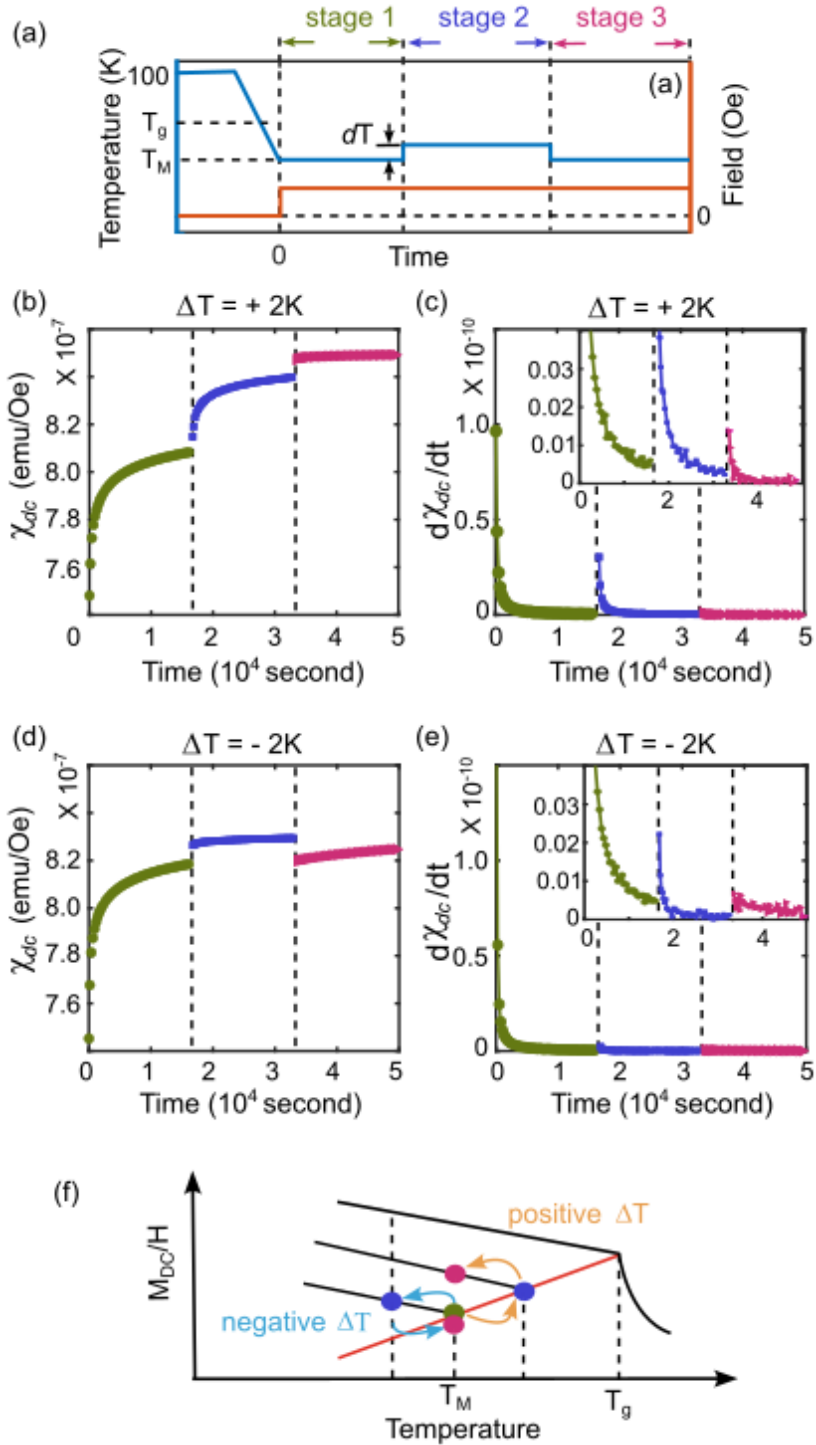

**Figure 6**. (a) Illustration of the field and temperature protocols for the measurements below. The three stages correspond to the time windows for aging, rejuvenation and memory effects. (b) and (c) The measured magnetic susceptibility with a positive temperature jump, $\Delta T = +2\,K$, as indicated in (a), and the corresponding magnetic fluctuations, defined as $d\chi_{dc}/dt$, as a function of time. (d) and (e) The measured magnetic susceptibility with a negative temperature change, $\Delta T = -2\,K$, and the corresponding $\Delta\chi_{dc}$. (f) Bifurcation between FC and ZFC curves occurring for different temperature and field protocols and its connection to the states within each thermal cycle.

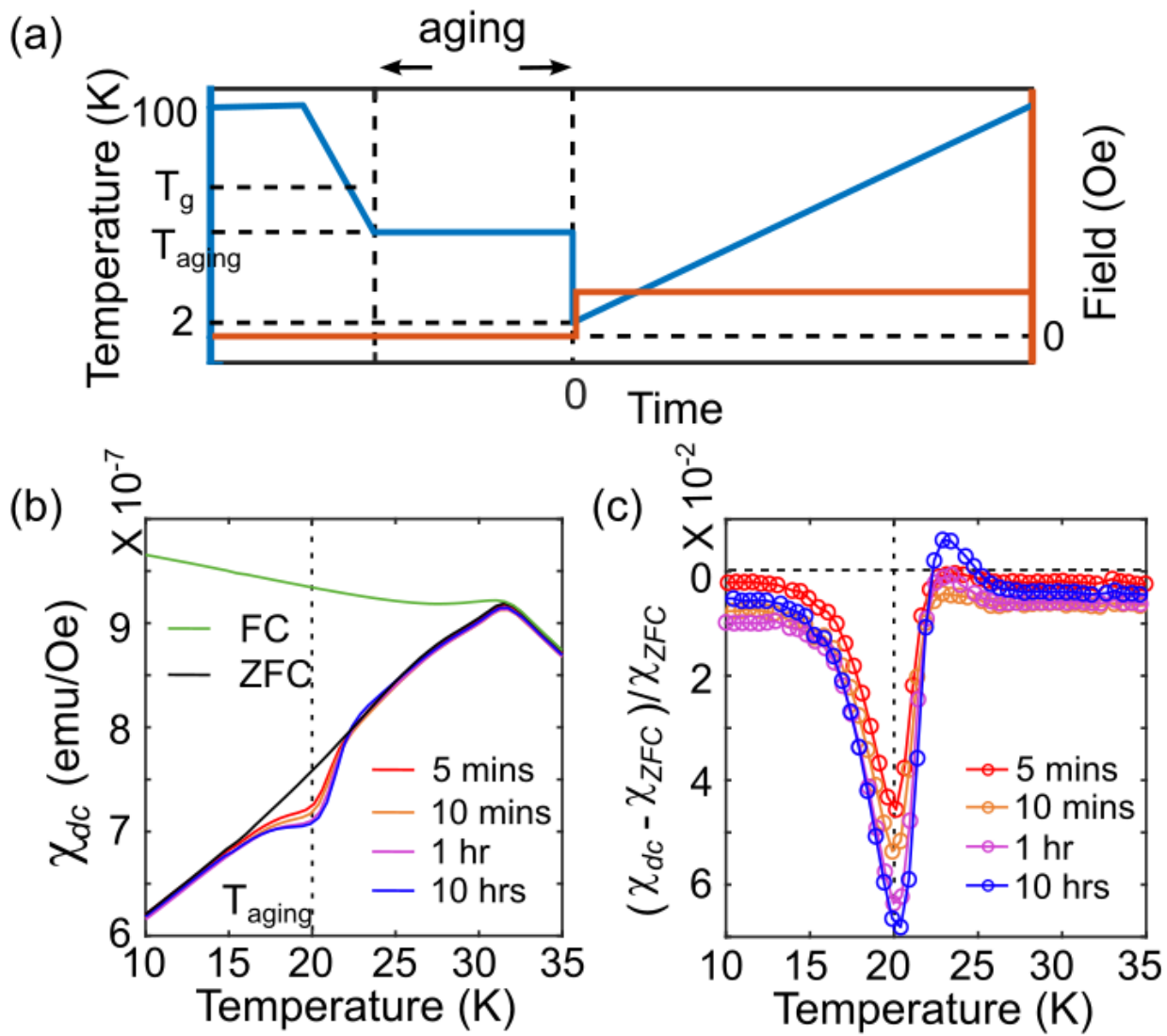

**Figure 7**. (a) Illustration of the field and temperature protocols for a step-cooling measurement. (b) The measured dc magnetic susceptibility as a function of temperature after FC, ZFC, and step-cooling procedures with selected aging times. (c) The corresponding difference in the magnetic susceptibility between step-cooling and ZFC data. The valley at 20 K indicates the memory effect.